\documentclass[aps,pre, groupedaddress, 12pt]{revtex4-2}
\usepackage[dvipsnames]{xcolor}
 \usepackage{graphicx}
 \usepackage{amsmath}
\usepackage{hyperref}
\usepackage[utf8]{inputenc}
 \usepackage{mathrsfs}
 \usepackage{amsmath}
\usepackage{amssymb}
\usepackage{amsmath}
 \usepackage{amssymb}
\newcommand{\be}{\begin{equation}}
\newcommand{\ee}{\end{equation}}
\newcommand{\ba}{\begin{align}}
\newcommand{\ea}{\end{align}}
\def\dbar{{\mathchar'26\mkern-12mu d}}
\begin{document}
\title{Microscopic Legendre Transform, Canonical
Ensemble and Jaynes' Maximum Entropy Principle}
\author{Ramandeep S. Johal}
 \email[e-mail: ]{rsjohal@iisermohali.ac.in \\
 https://orcid.org/0000-0002-9766-2814}
 \affiliation{ Department of Physical Sciences,
 Indian Institute of Science Education and Research Mohali,
 Sector 81, S.A.S. Nagar, Manauli PO 140306, Punjab, India}
\begin{abstract}
Legendre transform between thermodynamic quantities such as
the Helmholtz free energy and  entropy
plays a key role in the formulation of
the canonical ensemble. In the standard
treatment, the transform exchanges
the independent variable from 
the system's internal energy to its conjugate
variable---the inverse temperature of the heat reservoir.
In this article, we formulate a microscopic
version of the transform between the free energy and
Shannon entropy of the system, where
the conjugate variables are the microstate probabilities and
the energies (scaled by the inverse
temperature).
The present approach gives a non-conventional
perspective on the connection between
information-theoretic measure of entropy and
thermodynamic entropy.
We focus on the exact differential
property of Shannon entropy, utilizing it
to derive central relations within the canonical ensemble.
Thermodynamics of a system in contact with
the heat reservoir is discussed in this framework.
Other approaches, in particular,
Jaynes' maximum entropy principle
is compared with the present approach.
\end{abstract}
\maketitle
\newpage
\section{Introduction}
Legendre transform ($\mathscr{L}$) plays an important role in various
branches of physics. In mechanics, it connects Lagrangian
and Hamiltonian formalisms. In thermodynamics,
$\mathscr{L}$ serves to define
alternate quantities that
contain the same thermodynamic information
 as, say, the entropy of the system \cite{Callenbook}.
Such transformations are useful to describe systems
under different experimental conditions (see Appendix A
for example).
Statistical mechanics provides a microscopic 
underpinning for the equilibrium state within
the framework of ensemble theory 
\cite{Gibbs, Reifch6, Pathria}.
In canonical ensemble, the system energy is a 
random variable due to  
exchange of heat between the system and 
the heat reservoir. So,
the average energy of the system is defined  
over a probability distribution:
$U = \sum_{i=1}^{W} p_i \varepsilon_i$, 
where $\varepsilon_i \; (i=1,2,...,W)$ is
  the discrete energy eigenvalue 
  of the $i$th microstate of the system, populated with
  probability $p_i$ that satisfies
  the normalization condition 
  $\sum_{i=1}^{W} p_i =1$.
 Treating the composite 
 \textquotedblleft{system {\it plus} 
 reservoir}\textquotedblright 
  as an isolated system and invoking
Boltzmann's formula for thermodynamic entropy ($S = k_{\rm B}
\ln \Omega$), the probability distribution for the
system at thermal equilibrium is given by 
\be
 p_{i}^{*} = \frac{e^{-\beta^* \varepsilon_i}}
 {\sum_{i=1}^{W} e^{-\beta^* \varepsilon_i}},
 \label{pcan}
 \ee
which is the well-known Boltzmann or 
canonical distribution, with $\beta^*$ denoting
the inverse temperature of the reservoir. 
 We work in a system of units in which $k_{\rm B}=1$
so that entropy is a dimensionless quantity.
Then, temperature and energy are expressed
 in the same units. 
 The equilibrium free energy is obtained as
\be
\beta^* F (\beta^*) = -\ln \sum_{i=1}^{W} e^{-\beta^* \varepsilon_i}.
\label{bfstar}
\ee  
All
equilibrium properties of the system
may be calculated from the above free energy function. For example,
the average system energy is evaluated as
\be
U^* = \frac{\partial}{\partial \beta^*} (\beta^* F).
\label{ubf}
\ee
The equilibrium entropy, given by
$S(U^*) = \beta^* U^* - \beta^* F (\beta^*) $, can be cast
in terms of the canonical distribution, as
\be
S(U^*) = -\sum_{i=1}^{W} p_{i}^{*} 
\ln p_{i}^{*} \equiv \mathcal{S}^*.
\label{sgibbs}
\ee
The above expression is identical 
to the Shannon entropy \cite{Shannon} of a probability
distribution: $\mathcal{S}( p_1,...,p_W)
= -\sum_{i=1}^{W} p_{i} \ln p_{i}$, showing 
that thermodynamic entropy is
equal to the Shannon entropy ($\mathcal{S}^*$)  
of the equilibrium distribution.
 In fact, the formula for Shannon
entropy can also be derived directly from the Boltzmann
entropy by considering the multinomial multiplicity
of outcomes---in the limit of large numbers. 

The similarity of form between the
Shannon and the equilibrium entropy has been
a subject of much interest \cite{Hartley,Szilard,
Shannon,Wiener,Brill}  
in the foundations of statistical mechanics.
Jaynes \cite{Jaynes1957a, Jaynes1963} sought a fresh viewpoint
by which thermodynamic entropy and
 information-theoretic entropy could be looked upon
 as the same
 concept. In this pursuit, Jaynes came to regard
statistical mechanics as a  
problem of statistical inference applied to a system
with a limited prior information.  
 It was observed that
 thermodynamic entropy is the maximum of Shannon
entropy ${\cal S}( p_1,...,p_W)$,
obtained under the constraints
of a specified mean energy value
and normalization on probabilities.
 Jaynes' maximum entropy (Maxent) principle 
inspired applications in many diverse areas of science
and engineering \cite{Jaynes1979}.
Related progress into the role of information in physics has led to fundamental insights in the 
Maxwell's demon problem \cite{Leff_md}. 
Information is now 
regarded as a viable physical resource, and
thermodynamics of information processing is a thriving area of research \cite{Parrondo}.

The above connections between equilibrium
thermodynamic quantities on the one hand
and between entropic measures 
in statistical mechanics and 
information theory on the other, may be summarised
as follows. Equilibrium free energy is 
related to thermodynamic entropy via Legendre transform,
while thermodynamic entropy may be related to
Shannon entropy via Jaynes' principle. 
This raises a natural question:
Can the equilibrium free energy
be directly related to Shannon entropy
via an optimization procedure?
In this article, we analyze this  
relation in terms of a Legendre transform 
between these two quantities.
Clearly, with probabilities $p_i (i=1,...,W)$ as the apparent variables of Shannon entropy, we need to assign the conjugate variable for $p_i$.
 We term this procedure as
the microscopic Legendre transform
($\mathscr{L}_{\!\mathscr{M}}^{}$). 
Besides providing an alternate derivation 
of canonical
ensemble, this framework establishes a 
closer tie between thermodynamic and
Shannon entropies by underscoring
the exact differential property of  
entropy. 
Finally, the variational condition equivalent of the Maxent
method can be obtained within the present approach. The microscopic transform
thus forges an interesting connection with 
alternate approaches that derive the canonical distribution.
\section{microscopic Legendre transform
($\mathscr{L}_{\!\mathscr{M}}^{}$)}
Since $U$ is defined as an expectation value,
 the term $\beta^* U$  may be written as
$\sum_{i=1}^{W} p_i . (\beta^* \varepsilon_i)$,
suggesting  $\beta^* \varepsilon_i$ as the variable conjugate to $p_i$. At this stage, $\beta^*$ is just a 
parameter making
$\beta^* \varepsilon_i$ dimensionless---at par with
its conjugate variable $p_i$.
Thus, $1/\beta^*$ is some energy scale relevant to the problem
whose physical significance will be sought later.
With the values of $\beta^*$ and  $\varepsilon_i(i=1,...,W)$ as  specified,
we define $\mathscr{L}_{\!\mathscr{M}}^{}$
\cite{lt_su}:
\be
\beta^* F(\beta^* \varepsilon_1,\beta^* \varepsilon_2,...,\beta^* \varepsilon_W)
= \underset{p_1,p_2,...,p_W}{\rm Min}
\left\{
\sum_{i=1}^{W} p_i.(\beta^* \varepsilon_i) - \tilde{\mathcal{S}}(p_1,p_2..., p_W)
\right\},
\label{mltdef}
\ee
as the transform that replaces the set of variables
$(p_1,p_2,..., p_W)$ by the set
$(\beta^* \varepsilon_1,\beta^* \varepsilon_2,...,
\beta^* \varepsilon_W)$. We claim that $\mathscr{L}_{\!\mathscr{M}}^{}$ is the microscopic analog of the 
thermodynamic Legendre transform [Eq. (A.1)]. 
Here, the formulation in terms of dimensionless quantities 
$\tilde{\mathcal{S}}$ and $\beta^* F$ helps in
expressing the various relations in a symmetric form 
\cite{Zia}.

In the above, 
the entropy function $\tilde{\mathcal{S}}$ is
given by
\be
\tilde{\mathcal{S}}(p_1,p_2,..., p_W) =
- \sum_{i=1}^{W} p_i \ln p_i - \alpha \left(\sum_{i=1}^{W} p_i -1 \right),
\label{tils}
\ee
which explicitly specifies the constraint
of normalization. 
The parameter
$\alpha$ is the Lagrange multiplier accompanying 
the constraint, and is to be determined.
Clearly, the magnitude of 
$\tilde{\mathcal{S}}$ is equal to  
the corresponding Shannon entropy. 
The motivation to include the constraint 
in the definition of the entropy itself is due to  
the observation that the probabilities 
$p_i (i=1,2,...,W)$ do not constitute an independent set. Thereby, the partial derivatives of Shannon entropy---with respect to  $p_i$,
     are not defined \cite{comment_pds}. For the same reason,
     there is difficulty to define
     an exact differential of the entropy. 
     This point is important in order to 
     develop a statistical analog of thermodynamic
     entropy, since 
    in equilibrium thermodynamics, the 
    entropy of a system is a state function, and the difference in the entropy
between two nearby equilibrium states
is given by an exact differential. 
     Moreover, the definition of Legendre transform
          requires the variables
          in a set to be capable of independent variations.
Owing to the Lagrange multiplier method in Eq. (\ref{tils}), all the
probabilities can now be varied independently.

Thus, taking Eq. (\ref{tils}) as 
the statistical analog of thermodynamic entropy, 
we define its
exact differential as \cite{lm_alpha}
\be
d \tilde{\mathcal{S}} = \sum_{i=1}^{W}
\frac{\partial \tilde{\mathcal{S}} }{\partial p_i} dp_i,
\label{dsex}
\ee
where
\be
\frac{\partial \tilde{\mathcal{S}} }
{\partial p_i} = -(1+\ln p_i + \alpha).
\label{dsdpi}
\ee  
Consider first an isolated system
at equilibrium. Since the entropy
has maximal value, any process that 
the system undergoes spontaneously
must be a reversible process, implying
$d \tilde{\mathcal{S}} = 0$. In Eq. (\ref{dsex}), 
since the variations $dp_i$ are all independent, 
each partial derivative 
must vanish. From Eq. (\ref{dsdpi}), this yields
$p_i = e^{-(1+\alpha)} = 1/W$, i.e.   
an equiprobable distribution. This is 
the well-known  microcanonical ensemble
which in this case is equivalent to the Maxent principle.

Next, we consider the situation of a system in contact
with the heat reservoir. Suppose, 
as a result of the optimization 
in Eq. (\ref{mltdef}), 
we obtain the optimal distribution denoted by 
$p_i^* (i=1,2,...,W)$.
Then, the condition of the optimum is given as
\be
\left.\frac{\partial \tilde{\mathcal{S}} }{\partial p_i}
\right\vert_{p_i = p_{i}^{*}} =
\beta^* \varepsilon_i.
\label{dspep}
\ee
As is the purpose of a Legendre transform, 
$\beta^* F$ encodes the same information
as the function $\tilde{\mathcal{S}}$, but 
expressed in terms of the derivatives
$\partial \tilde{\mathcal{S}} /{\partial p_i}$,
instead of $p_i$. 

Combining Eqs. (\ref{dsdpi}) and (\ref{dspep}),
we obtain:
\be 
\beta^* \varepsilon_i +  1+\ln p_i^* + \alpha = 0.
\label{eqmco}
\ee
Note that Eq. (\ref{dsdpi}) is due to the
requirement of an exact differential, whereas
Eq. (\ref{dspep}) is implicit in the 
definition of $\mathscr{L}_{\!\mathscr{M}}^{}$.
Solving Eq. (\ref{eqmco}) for $p_i^*$, we obtain
$p_i^* = e^{-(1+\alpha)} e^{-\beta^* \varepsilon_i}$.
Summing over all states and using the normalization constraint,
we get: $(1+\alpha) = \ln \sum_{i=1}^{W} e^{-\beta^* \varepsilon_i}$, 
and so we obtain $p_i^* (i=1,2,...,W)$ in the form
of Eq. (\ref{pcan}).

It may be remarked that the above procedure
bears some semblence to the Maxent method which 
also optimizes Shannon entropy (subject to 
mean-value constraints). On the other hand, the microscopic transform 
here is defined with a specified parameter $\beta^*$,
but unlike the transform in thermodynamics [Eq. (A.1)],
the optimization is performed over the space of probability 
distributions. The mean energy is then determined only
after the optimal distribution is obtained. 

Using Eq. (\ref{tils}) with $p_i = p_i^*$, the entropy at the
stationary point is given by
\begin{align}
 \tilde{\mathcal{S}}^* \equiv 
 \tilde{\mathcal{S}}(p_1^*, p_2^*,..., p_W^*) =
 \sum_{i=1} ^{W} p_i^*.(\beta^* \varepsilon_i)
  +\ln \sum_{i=1}^{W} e^{-\beta^* \varepsilon_i},
\end{align} 
which is equal to the thermodynamic entropy [Eq. (\ref{sgibbs})]. 
Then, the Legendre transform $\mathscr{L}_{\!\mathscr{M}}^{}$:
\begin{align}
\beta^* F(\beta^* \varepsilon_1,\beta^* \varepsilon_2,...,\beta^* \varepsilon_W) & = 
\sum_{i=1} ^{W} p_i^*.(\beta^* \varepsilon_i) -
\tilde{\mathcal{S}}(p_1^*, p_2^*,..., p_W^*), 
\label{mlt}
\end{align}
yields the standard expression for 
the equilibrium free energy :
$\beta^* F
= -\ln \sum_{i=1}^{W} e^{-\beta^* \varepsilon_i}$.

The structure of $\mathscr{L}_{\!\mathscr{M}}^{}$ 
ensures that $\beta^* F$ 
depends only upon the set
$(\beta^* \varepsilon_1,\beta^* \varepsilon_2,...,
\beta^* \varepsilon_W)$.
Denoting the function to be minimized 
in Eq. (\ref{mltdef}) as $\beta^* \mathcal{F}$,  
consider an infinitesimal variation: 
\begin{align}
 d(\beta^* \mathcal{F}) &= \sum_{i=1}^{W} p_i. d(\beta^*\varepsilon_i)
              + \sum_{i=1}^{W} \beta^*\varepsilon_i. dp_i
              +\sum_{i=1}^{W} (1 + \ln p_i) dp_i
              + \alpha \sum_{i=1}^{W} dp_i \nonumber \\
            & =   \sum_{i=1}^{W} p_i d(\beta^*\varepsilon_i)
              + \sum_{i=1}^{W} (\beta^*\varepsilon_i
              + 1 + \ln p_i + \alpha) dp_i.
              \label{dbfall}
\end{align}
Due to the condition (\ref{eqmco}) signifying 
the stationary point ($p_i = p_i^*$),
the second sum above vanishes and 
we obtain the exact differential:
\be
d(\beta^* {F}) = \sum_{i=1}^{W} p_i^* d(\beta^*\varepsilon_i),
\label{bfexact}
\ee
where 
\begin{equation}
p_i^* = \frac{\partial (\beta^* {F})}{ \partial
(\beta^* \varepsilon_i)},
\label{dbfde}
\end{equation}
which may be easily verified by using the 
explicit expression for $\beta^* {F}$.
For the inverse transform of $\mathscr{L}_{\!\mathscr{M}}^{}$ and other relations, 
 see Appendix C.

 Finally, we note that the parameter
 $\beta^*$ can be interpreted as the inverse temperature of
 the system at equilibrium, defined as
 \begin{equation}
  \left. \frac{\partial \tilde{\mathcal{S}}}{\partial U}
  \right\vert_{p_i = p_{i}^{*}}
= \left. \frac{\partial \tilde{\mathcal{S}}}{\partial p_i}
\frac{\partial p_i}{\partial U} \right\vert_{p_i = p_{i}^{*}}
=\beta^*,
  \end{equation}
where Eq. (\ref{dspep}) is used along with ${\partial U}/{\partial p_i} = \varepsilon_i$. In the following section, 
the (inverse) temperature of the system is equated to
that of the equilibrating reservoir.
\section{Thermodynamic perspective}
We now focus on the thermodynamics of
a system in contact with the heat reservoir.
The second law stipulates that the total
entropy of the system and the reservoir attains
the maximum value at equilibrium, whereby
the temperature of the system is equal to that
of the reservoir.
Then, a quasi-static, infinitesimal process involving an
exchange of heat between the system and the
reservoir must
be a reversible process, implying
\begin{equation}
d \tilde{\mathcal{S}}^* + d S_{\rm R} = 0.
 \label{dsreva}
\end{equation}
Now, suppose the amount of heat
added to the system is $\;\dbar Q^*$ so that 
the corresponding amount for the reservoir is
$- \dbar Q^*$. As explained above, 
the specified parameter $\beta^*$ can be interpreted as
the inverse temperature of system and hence of the associated heat reservoir.
The definition of a heat reservoir implies
 $dS_{\rm R} = -\beta^* \;\dbar Q^*$ and so Eq. (\ref{dsreva}) yields
$d \tilde{\mathcal{S}}^* = \beta^* \; \dbar Q^*$.
Now, from Eq. (\ref{mlt}), we can write
\be
d(\beta^* {F}) = \sum_{i=1}^{W} p_i^* d(\beta^*\varepsilon_i)
              + \beta^* \sum_{i=1}^{W} \varepsilon_i  dp_i^*
              -d \tilde{\mathcal{S}}^*.
\label{dbflt}
\ee
Using Eq. (\ref{bfexact}), the above equation simplifies to
$d \tilde{\mathcal{S}}^*
= \beta^* \sum_{i=1}^{W} \varepsilon_i dp_i^*$.
Therefore, the infinitesimal heat exchanged
in the reversible process involving the system and the reservoir
is $\dbar Q^* = \sum_{i=1}^{W} \varepsilon_i dp_i^*$,
which agrees with the standard statistical definition of
quasi-static heat in an infinitesimal process
\cite{Reifch6}.
%
Alternately, and more directly, we can combine the exact differential of $\tilde{\mathcal{S}}$ [Eq. (\ref{dsex})]
with Eq. (\ref{dspep}), and so
obtain the same expression for the heat exchanged, as above.

Further, due to Eq. (\ref{bfexact}),
an infinitesimal change in $\beta^* F$ 
is equal to the equilibrium average of the differential changes
in $\beta^*\varepsilon_i$. Now, a thermodynamic
system is subject to control of some
macroscopic parameters such as the reservoir
temperature or the volume of the system.
Considering that the energy eigenvalues are
a function of the volume (and possibly
other such parameters which are held constant),
we can express the variation in
$\beta^*\varepsilon_i$ as follows.
 \begin{align}
 d(\beta^*\varepsilon_i) & = \varepsilon_i d\beta^* + \beta^* d \varepsilon_i, \\
                       & =  \varepsilon_i d\beta^* +
                       \beta^* \frac{\partial \varepsilon_i}{\partial V} dV.
\end{align}
Substituting the above in Eq. (\ref{bfexact}) and using
the definition of average pressure \cite{Reifch6}:
\be
P = -\sum_{i=1}^{W} p_i^*  \frac{\partial \varepsilon_i}{\partial V},
\ee
we recover the thermodynamic relation:
\be
d (\beta^* F) = U^* d \beta^* - \beta^* P dV.
\label{dbfth}
\ee
This derivation shows that 
the relation (\ref{bfexact}) is equivalent to the
above thermodynamic expression which is usually written  
in terms of the macroscopic variables.
As a special case, if the parameter $\beta^*$
 is held fixed, we conclude that
 $d F = -P dV = \;\dbar W$, implying that the  work done
 in a reversible process is
 equal to the change in the Helmholtz free energy
 of the system in thermal
 equilibrium with a given reservoir.

As mentioned earlier, entropy
is regarded as a state function in macroscopic thermodynamics.
The change in the entropy of the system for
a  quasi-static, infinitesimal process connecting
two equilibrium states can be
expressed as an exact differential \cite{Callenbook}:
\be
dS = \beta^*{dU} + \beta^* P dV,
\label{dsth}
\ee
where $U$ and $V$ are the independent variables,
assuming $N$ to be fixed.
According to the first
law of thermodynamics, $dU=\dbar Q + \dbar W$.
Thus, Eq. (\ref{dsth}) gets simplified to
$dS = \beta^* \dbar Q$.

Likewise, within the statistical framework too, an infinitesimal
change in the equilibrium mean energy during a 
quasi-static process 
can be split as
\begin{align}
dU^* & = \sum_{i=1} ^{W} \varepsilon_i dp_i^*
         + \sum_{i=1} ^{W} p_i^* d\varepsilon_i.
         \label{dusplit}
    \end{align}
We have already argued that 
the heat exchanged in the process is
$\;\dbar Q^* = \sum_{i=1} ^{W} \varepsilon_i dp_i^*$
and so the work is identified with $\;\dbar W^* = \sum_{i=1} ^{W} p_i^* d\varepsilon_i$.
%
\section{Alternate derivations of canonical distribution}
Finally, we highlight that the canonical distribution may also be 
inferred from the condition of a reversible
process between the system and the reservoir.
Unlike the standard treatment \cite{Callenbook,Pathria} based on Boltzmann entropy, we make use of Shannon entropy
to arrive at the desired variational condition. 
As discussed above, a quasi-static, infinitesimal
process at equilibrium is a reversible process.
Suppose that we keep the energy eigenvalues fixed so that 
no work is performed. Then,
from the first law,  $\;\dbar Q = dU =
\sum_{i=1}^{W} \varepsilon_i dp_i$.
Since this heat is exchanged with the heat reservoir,
so the change in the entropy of the latter is
 $dS_{\rm R} = - \beta^* \;\dbar Q$.
 On the other hand, from Eq. (\ref{tils}),
the change in the entropy of the system
is given by the exact differential:
$ d \tilde{\mathcal{S}}
 = - \sum_{i=1}^{W} (1 + \ln p_i + \alpha)dp_i$.
So, the condition for a reversible process,
$d \tilde{\mathcal{S}} + d S_{\rm R} = 0$,
yields
\be
- \sum_{i=1}^{W} (1 + \ln p_i + \alpha)dp_i -
\beta^* \sum_{i=1}^{W}   \varepsilon_i dp_i  = 0,
\label{dsdp}
\ee
which may be expressed in the form 
\begin{equation}
               \sum_{i=1}^{W} (\beta^*\varepsilon_i
              + 1 + \ln p_i + \alpha) dp_i =0.
              \label{dpo}
\end{equation}
Note that $\beta^*$ is already specified as
the reservoir temperature. 
Now, as the variations $dp_i$ are independent,
the above condition implies that the coefficient
of each $dp_i$ vanishes, thus yielding
the condition equivalent to Eq. (\ref{eqmco}),
 and so the canonical distribution.
Note that there is a subtle difference
of viewpoint in the reversibility 
condition used in Eq. (\ref{dsreva}).
There, we knew the equilibrium distribution 
(denoted by $p^*$), but here, we are 
deriving the same {\it from} the reversibility
condition.

Eq. (\ref{dpo}) is
similar to conditions obtained in other derivations
of canonical distribution, in particular,
the method of most probable distribution \cite{Wallis, Reifch6}
and the maximum
entropy principle of Jaynes \cite{Jaynes1957a}.
Both these methods maximize a quantity subject 
to the constraints of normalization and a specified mean
energy, and make use of the method of Lagrange multipliers.
The so-called Wallis method \cite{Wallis} maximises a
combinatorial quantity. Further, it defines
the notion of probability
in the frequency sense, whereby the canonical 
distribution is recovered in the limit of large numbers.
On the other hand, Jaynes regards probabilities
in epistemic or subjective sense.
In particular, Jaynes maximized
 $\mathcal{S} = -\sum_{i=1}^{W} p_i \ln p_i$,
 taken as a measure of uncertainty of the observer
 regarding the actual state of the system,
under the given  constraints
$\sum_{i=1}^{W} p_i = 1$ and 
$\sum_{i=1}^{W} p_i \varepsilon_i = U$.
For each constraint, a distinct Lagrange multiplier
is introduced. The target function to be optimized
may be written as follows.
\begin{equation}
 \mathscr{S} =  -\sum_{i=1}^{W} p_i \ln p_i
               - \alpha \left(\sum_{i=1}^{W} p_i -1\right)
               - \beta \left(\sum_{i=1}^{W} p_i \varepsilon_i -U\right),
               \label{mep}
\end{equation}
where $\alpha$ and $\beta$ both are Lagrange
multipliers which are determined by satisfying the given constraints.
The optimization, $d\mathscr{S}=0$, then yields
a condition equivalent to Eq. (\ref{dpo}).
Recently, other derivations of the canonical 
distribution have appeared in literature as an 
alternative to the Maxent approach (see for example,
Refs. \cite{Plastino2005, Bagci2012}).

\section{Conclusions}
For a thermodynamic system in equilibrium
with a heat reservoir,
the entropy and Helmholtz free energy are related
to each other by Legendre transform.
In this mathematical structure, 
the internal energy $U$ and the 
inverse reservoir temperature $\beta^*$ play
the role of conjugate variables.
This relation is also maintained
within the standard treatment of canonical ensemble.
We have studied an alternate formulation
of this transform between Shannon entropy
and Helmholtz free energy. Here,
the variables of entropy are
the microstate probabilities whose
conjugate variables are the microstate
energies scaled by $\beta^*$. 
In this construction, an essential role
is played by the exact differential 
property of entropy which is equivalent to 
defining the probabilities in a distribution
as independent variables.  
This property then allows the analysis of the 
equilibrium condition in a thermal contact, based on  
the Shannon form of entropy.


The proposed transform is based on a different
premise than the 
Maxent procedure of Jaynes, although the final 
equilibrium state is predicted to be the same. The latter method maximizes 
a measure of uncertainty (Shannon entropy)
under the specified mean values. 
Jaynes was led
to conclude that statistical mechanics
may be dissociated from physical arguments 
and advocated to view it instead
as a problem in statistical inference. 
As is well known, the original motivation for
the theory of statistical mechanics
was to justify thermodynamics which earlier
had an empirical basis.
But, in veaning statistical mechanics from a
physical basis, its thermodynamic
relevance might seem remote.
The microscopic transform [Eq. (\ref{mltdef})] is based
on an optimization over the probability distributions, 
for  given values of parameter $\beta^*$ and the energy eigenvalues. This parameter can be interpreted
as the inverse temperature of the system at the optimal 
distribution. In the context of system-reservoir
contact, $\beta^*$ gets identified with the 
inverse temperature of the reservoir. 
For a reversible process at equilibrium, 
the thermal contact scenario also yields the 
variational condition equivalent to Maxent.  
Thus, by formulating the exact differential
property of Shannon entropy,  
we have generalized the usual system-reservoir approach
in which the entropy is defined by the Boltzmann formula.

 The mathematical formulation can be 
 easily generalized to other ensembles, such 
 as the grand canonical ensemble. We have 
 restricted to the case of discrete state space,
 and so a generalization to continuous variables 
 needs to be revisited. Similarly, extensions
 to quantum density matrices and irreversible
 processes are important lines of inquiry.
 Because of the ubiquity of Maxent methods and 
 concepts of statistical manifolds in other disciplines such as 
 dynamical systems \cite{Beck}, machine learning \cite{Mackay, Simeone} and 
 information geometry \cite{Amari,Nielsen}, 
 the transform can potentially be adapted to deal with various
 `Hamiltonians' and measures of
 uncertainty. Besides these possible lines of
 research, the proposed microscopic transform 
 can be a useful technique in physics pedagogy as well.
\begin{acknowledgments}
The author acknowledges useful discussions and comments
from Sumiyoshi Abe, Armen Allahverdyan, Deepak Dhar and Henni Ouerdane.
\end{acknowledgments}
\section*{Funding Declaration}
The research work described in this paper received no funding.

\section{Appendix}
\subsection{Legendre transform in thermodynamics}
Helmholtz free energy ($F$) of a
system having the volume $V$,
the number of particles $N$, and in 
thermal equilibrium with
a heat reservoir at a specified inverse temperature $\beta^* = 1/T^*$, may be defined
as the  $\mathscr{L}$ transform of the entropy as \cite{lt_su, comment_Massieu}:
\begin{equation}
 \beta^* F(\beta^*,V,N) =  \underset{U}{\rm Min} \left\{
 \beta^* U - S(U,V,N)
 \right\}.
 \tag{A.1}
 \label{bfth}
\end{equation}

Now, suppose the above optimization yields 
a stationary point at $U=U^*$, implying 
\be
\beta^* = \left.\frac{\partial S}{\partial U} \right\vert_{U^*}
\tag{A.2}
\label{theqb}
\ee
i.e. at the energy $U^*$, the system 
inverse temperature (${\partial S}/{\partial U} \equiv
\beta$) is equal to that of the reservoir, or,
in other words, 
the system is in thermal equilibrium with the reservoir.
Upon second variation, we get $\partial^2 S/\partial U^2 \vert_{U^*} < 0$,
which follows due to concavity of the entropy $S(U)$, 
and this implies a minimum for the quantity
$\{\beta^* U- S\}$ as a function of $U$.

Eq. (\ref{theqb}) can be used to write $U^*$
as a function of $\beta^*$, denoted by 
$U^*(\beta^*)$.
Thus, we can express Eq. (\ref{bfth}) as follows.
\be
\beta^* F(\beta^*) = \beta^* U^*(\beta^*) - S(U^*(\beta^*)).
\tag{A.3}
\ee
 where we have suppressed the passive variables ($V,N$) for brevity.

Likewise, $S(U^*)$ can be defined as the inverse
Legendre transform of the free energy. For a specified parameter
$U^*$, we have:
\be
S(U^*) =  \underset{\beta}{\rm Min} \{\beta U^* - \beta F(\beta) \}.
\tag{A.4}
\ee
whose stationary point 
\be
U^* - \frac{\partial}{\partial \beta} (\beta F) = 0,
\tag{A.5}
\ee
is obtained at $\beta = \beta^*$.
The equilibrium entropy of
the system can be written as: 
\be
S(U^*) = \beta^*(U^*) U^* - \beta^*(U^*) F(\beta^*(U^*)).
\tag{A.6}
\ee
\subsection{Legendre transform in statistical mechanics}
We summarise how the Legendre transform
structure described above is carried over
in the canonical ensemble. The free 
energy function is now defined via relation
\be
\beta F (\beta) = -\ln \sum_{i=1}^{W} e^{-\beta \varepsilon_i}.
\tag{B.1}
\ee
Using this definition in Eq. (A.4),   
we obtain:
\be
S(U^*) =  \underset{\beta}{\rm Min} \{\beta U^* + \ln \sum_{i=1}^{W} e^{-\beta \varepsilon_i}\},
\tag{B.2}
\ee
whose stationary point yields: 
\be
U^* 
 = \frac{\sum_{i=1} ^{W} \varepsilon_i e^{-\beta^* \varepsilon_i}}
 {\sum_{i=1}^{W} e^{-\beta^* \varepsilon_i}}, 
 \tag{B.3}
 \ee
 from which we may determine 
 $\beta^*$ corresponding to a specified value $U^*$.
Further, since the average energy is defined as   
$U^*= \sum_{i=1} ^{W} \varepsilon_i p_{i}^{*}$,
Eq. (B.3) yields the optimal
probability distribution corresponding to
the stationary
point, as
 \be
 p_{i}^{*} = \frac{e^{-\beta^* \varepsilon_i}}
 {\sum_{i=1}^{W} e^{-\beta^* \varepsilon_i}}.
 \tag{B.4}
 \ee
Thence, the equilibrium free 
energy is given by
\be
\beta^* F (\beta^*) = -\ln \sum_{i=1}^{W} e^{-\beta^* \varepsilon_i}.
\tag{B.5}
\ee
From the knowledge of the reservoir temperature
and the microstate energies of the system, all
equilibrium properties of the system
may be calculated. For example,
$U^* = ({\partial}/{\partial \beta^*}) (\beta^* F)$,
which is as well a consequence of the
Legendre transform [Eq. (A.4)] discussed above.  
  
In terms of the equilibrium distribution,
the equilibrium entropy $S(U^*)$ can be cast in the form:
$S(U^*) = -\sum_{i=1}^{W} p_{i}^{*} 
\ln p_{i}^{*} \equiv S^*$.
The form of the expression is identical 
to Shannon entropy of a probability
distribution, $\mathcal{S} = -\sum_{i=1}^{W} p_{i} 
\ln p_{i}$. However, note that 
whereas thermodynamic entropy $S^*$ above is
a concave function of the equilibrium mean energy 
$U^*$, Shannon entropy is a concave function
of the given probability 
distribution \cite{Portamana}. 
\subsection{Some relations for $\mathscr{L}_{\!\mathscr{M}}^{}$ and the inverse microscopic transform}
We have seen that 
\begin{equation}
p_i^* = \frac{\partial (\beta^* {F})}{ \partial
(\beta^* \varepsilon_i)}.
\label{dbfde}
\tag{C.1}
\end{equation}
An interesting relation follows, as
\be
\sum_{i=1}^{W}
\frac{\partial (\beta^* {F})}{ \partial
(\beta^* \varepsilon_i)} =1.
\tag{C.2}
\ee
Further, 
the equilibrium energy may alternately be
given  by the formula:
\be
U^* = \frac{1}{\beta^*}
\sum_{i=1}^{W}\left. p_i
\frac{\partial \tilde{\mathcal{S}} }{\partial p_i}
\right\vert_{p_i = p_{i}^{*}}.
\tag{C.3}
\ee
For the inverse microscopic transform
corresponding to a given probability distribution $p_i^*$, we can write
\be
\tilde{\mathcal{S}}(p_1^*, p_2^*,..., p_W^*) 
=  \underset{\beta \varepsilon_1, \beta \varepsilon_2, ...,\beta \varepsilon_W}{\rm Min} \left \{ \sum_{i=1}^{W} 
(\beta \varepsilon_i) p_i^* + \ln \sum_{i=1}^{W} e^{-\beta \varepsilon_i} \right \},
\tag{C.4}
\ee
whose stationary point implies that 
the given distribution is in the canonical
form with respect to the optimal set 
of $(\beta \varepsilon_1, \beta \varepsilon_2, 
...,\beta \varepsilon_W)^*$.
Note that if the energies are specified, then 
this procedure is equivalent to Eq. (B.2) which 
yields $\beta = \beta^*$. Within the microscopic
transform, we  treat 
$\beta \varepsilon_i$ as a collective variable. 
%

%

\begin{thebibliography}{30}
\bibitem{lt_su} Legendre transform of a concave function
$F(x_1,...,x_W)$, with $x_i (i=1,2,...,W)$ as
{\it independent} variables, and the
specified set of values 
$(y_1,y_2,...,y_W)$, 
is defined by the function:
\[
G(y_1,y_2,...,y_W) = 
\underset{x_1,x_2,...,x_W}{\rm Min}
\left\{
\sum_{i=1}^{W} x_i y_i - F(x_1,x_2,...,x_W)
\right\},
\]
where the stationary point is given by
the condition: 
\[
 y_i = \frac{\partial F}{\partial x_i}, 
 \quad \forall i=1,2,...,W
\]
In the context of Eq. (\ref{mltdef}), 
$F(x_1,x_2,...,x_W) \equiv \tilde{\mathcal{S}}(p_1, p_2,..., p_W)$
is a concave function of the probability distribution.
\bibitem{comment_pds}
One may choose to eliminate one of the probabilities and
deal with the remaining $(W-1)$ probabilities as
independent variables.
However, owing to non-unique choice for this independent
set, the method is arbitary. As the constraint
 ($\sum_{i=1}^{W} p_i =1$) 
involves all the probabilities in a symmetric fashion, so the method of Lagrange multipliers is one method that treats all the variables in a symmetric manner.
\bibitem{lm_alpha}
More rigorously, $ \tilde{\mathcal{S}}$ 
can be additionally regarded  as function of
the parameter $\alpha$. So, in general,
the exact differential (\ref{dsex}) may be written with 
an extra term: $(\partial \tilde{\mathcal{S}} / \partial \alpha) d \alpha$.
Since, $\partial \tilde{\mathcal{S}} / \partial \alpha = 
1-\sum_{i=1}^{W} p_i=0$, so this term drops out.
\bibitem{dqinexact}
The fact that $\;\dbar Q = \sum_{i=1} ^{W} \varepsilon_i dp_i^*$
is inexact or process-dependent may be seen as follows. For a process connecting
two specified equilibrium states with a difference in energy 
$dU$, if we keep the energy levels fixed, then no work is 
performed and $ \;\dbar Q = dU$. On the other hand, 
we may convert $dU$ to work, or vice versa, by varying
the energies while keeping the probabilities fixed,
in which case
$\;\dbar Q  = 0$.
\bibitem{comment_Massieu} More precisely, 
Legendre transforms of entropy are also known as
Massieu functions or Massieu-Planck functions, see 
e.g. Ref. \cite{Callenbook}, Chap. 5, p. 151.
\end{thebibliography}
\end{document}